\documentclass[aps,pra,twocolumn,superscriptaddress]{revtex4-1}

\usepackage{graphicx}
\usepackage{hyperref}
\usepackage{amsmath}
\usepackage{amssymb}
\usepackage{epstopdf}
\usepackage{multirow}
\usepackage{threeparttable}
\usepackage{array}
\usepackage{booktabs}
\usepackage{xcolor}
\usepackage{mathtools}
\usepackage{ulem}

\begin{document}

\title{Mid-infrared snapshot spectral imaging via nonlinear radial dispersion}
\author{Jianan Fang}
\affiliation{State Key Laboratory of Precision Spectroscopy, and Hainan Institute, East China Normal University, Shanghai 200062, China}
\affiliation{Chongqing Key Laboratory of Precision Optics, Chongqing Institute of East China Normal University, Chongqing 401121, China}

\author{Kun Huang}
\email{khuang@lps.ecnu.edu.cn}
\affiliation{State Key Laboratory of Precision Spectroscopy, and Hainan Institute, East China Normal University, Shanghai 200062, China}
\affiliation{Chongqing Key Laboratory of Precision Optics, Chongqing Institute of East China Normal University, Chongqing 401121, China}
\affiliation{Collaborative Innovation Center of Extreme Optics, Shanxi University, Taiyuan, Shanxi 030006, China}

\author{Ruiyang Qin}
\affiliation{State Key Laboratory of Precision Spectroscopy, and Hainan Institute, East China Normal University, Shanghai 200062, China}

\author{Jixi Zhang}
\affiliation{State Key Laboratory of Precision Spectroscopy, and Hainan Institute, East China Normal University, Shanghai 200062, China}

\author{Heping Zeng}
\email{hpzeng@phy.ecnu.edu.cn}
\affiliation{State Key Laboratory of Precision Spectroscopy, and Hainan Institute, East China Normal University, Shanghai 200062, China}
\affiliation{Chongqing Key Laboratory of Precision Optics, Chongqing Institute of East China Normal University, Chongqing 401121, China}
\affiliation{Shanghai Research Center for Quantum Sciences, Shanghai 201315, China}
\affiliation{Chongqing Institute for Brain and Intelligence, Guangyang Bay Laboratory, Chongqing, 400064, China}

\begin{abstract}
Mid-infrared (MIR) spectral imaging provides chemically specific contrast through molecular vibrational fingerprints, yet snapshot acquisition remains severely limited by the lack of high-sensitivity detectors and efficient spectral encoding mechanisms. Here we introduce snapshot MIR spectral imaging based on intrinsic nonlinear radial dispersion, in which wavelength-dependent phase matching simultaneously enables frequency upconversion and spectral multiplexing. Different spectral components are mapped to distinct output angles within a 4$f$ imaging architecture, enabling single-shot spectral encoding without external dispersive elements. In combination with speckle illumination encoding, spectral information is compressed and recovered without additional coding components. Leveraging nonlinear upconversion to the visible, the approach achieves room-temperature MIR spectral imaging with sensitivity approaching 1 photon/pixel/pulse across a broad spectral range from 2.5 to 4.0 $\mu$m. This work transforms spectral encoding from an external optical function into an inherent property of the nonlinear imaging process, providing a general route to high-sensitivity snapshot MIR spectral imaging.
\end{abstract}


\maketitle

\vspace{8pt}
\noindent{\fontfamily{phv}\selectfont 
	\textbf{INTRODUCTION}}
\vspace{4pt}
\newline
Spectral imaging enables the simultaneous acquisition of spatial and spectral information and has become a powerful, non-invasive tool for chemical and material analysis \cite{Khan2018IEEE, Gao2015JB}. In particular, mid-infrared (MIR) spectral imaging is of special interest because this spectral region hosts fundamental molecular vibrational resonances that provide chemically specific contrast \cite{Vodopyanov2020Book}. By correlating spatial morphology with spectral composition, MIR spectral imaging has found broad applications in biomedical diagnostics, materials characterization and environmental monitoring \cite{Rodrigues2022ABC, Hermes2018JO}. As the benchmark technique for chemical imaging, Fourier-transform infrared (FTIR) spectroscopy achieves high spectral resolution through mechanical scanning of an interferometer arm \cite{Griffiths1983Science, Dorling2013TB}. To alleviate the speed limitations imposed by mechanical motion, electrically tunable approaches have been developed, such as rapid wavelength tuning of illumination sources \cite{Yeh2015AC} and high-speed spectral filtering of broadband signals \cite{Ward2015JPCS, Wang2020JMM}. Nevertheless, these methods fundamentally adhere to Nyquist-sampling-based acquisition paradigms, in which spatio-spectral information is acquired sequentially. As a result, the measurement time and data volume scale linearly with the desired spatial or spectral resolution, rendering such approaches ill-suited for capturing dynamic scenes and imposing substantial burdens on data storage and post-processing \cite{Baier2020PCI, Hiramatsu2019SA}.

Snapshot spectral imaging has therefore emerged as an attractive alternative, aiming to acquire spectral information within a single exposure \cite{Hu2024AP, Huang2022Light}. Early snapshot approaches, including image-mapping spectroscopy \cite{Liu2022AO} and light-field imaging spectroscopy \cite{Su2015Optik}, project slices of the spectral datacube onto a two-dimensional detector and reconstruct the full datacube computationally. While conceptually straightforward, these methods typically trade spatial resolution for spectral bandwidth. More recently, advances in computational imaging and compressed sensing have enabled more efficient information recovery beyond Nyquist limits \cite{Bian2024Nature, Dong2025AP, Yao2025Nature, Donoho2006TIT, Park2024PhotoniX}. Within this framework, spectral compressive imaging techniques encode three-dimensional spectral datacubes into two-dimensional measurements using spatial or spectral modulation, followed by algorithmic reconstruction \cite{Wagadarikar2008AO, Cao2011TPAMI, Cao2016SPM}. Among these, coded-aperture snapshot spectral imaging (CASSI) employs a fixed mask and a dispersive element to perform wavelength-dependent multiplexing in a single shot \cite{Gehm2007OE}. This architecture offers high optical throughput and compact implementations \cite{Xu2023TM, Zhang2023NC}, with reconstruction achieved via regularized optimization or deep learning algorithms \cite{Yuan2016ICIP, Boyd2011FTML, Barbastathis2019Optica, Choi2017ACM}. Inspired by diverse spectral responses of optical media such as scatterers \cite{Li2019Optica, Monakhova2020Optica}, metasurfaces \cite{Hua2022NC, McClung2020SA} and diffractive optical elements \cite{Majumder2025Optica, Arguello2021Optica, Hu2024IPT}, a range of alternative snapshot architectures has since been proposed.

However, extending snapshot spectral imaging into the MIR regime remains particularly challenging. State-of-the-art MIR detectors based on microbolometers or narrow-bandgap semiconductors typically suffer from high dark noise, limited pixel density and slow readout speeds \cite{Razeghi2014PR, Wang2019Small}. Recently, two-photon absorption has enabled MIR spectral imaging on a Si-based chip \cite{Knez2020LSA, Knez2022SA} albeit with limited detection efficiency due to the high-order nonlinearity. In parallel, efficient spectral encoding components, such as coded apertures and dispersive elements, are difficult to implement in the MIR owing to fabrication constraints and long-wavelength diffraction effects \cite{Wu2019AO}. Although digital micromirror devices offer programmable modulation, their performance in the MIR is severely compromised, necessitating complex calibration and correction procedures \cite{Wu2020OE}. Together, the lack of high-sensitivity detectors and scalable spectral encoding mechanisms has largely confined practical snapshot implementations to the visible and near-infrared regimes.

Frequency upconversion imaging provides an indirect yet powerful route to overcoming these limitations \cite{Barh2019AOP, Dam2012NP, Paterova2020SA, Kviatkovsky2020SA, Mrejen2020LPR, Ge2023APN}. By converting MIR photons to shorter wavelengths via nonlinear optical processes, upconversion imaging enables the use of mature silicon-based detectors, achieving room-temperature MIR imaging with single-photon sensitivity and high readout speeds \cite{Huang2022NC}. Building on this principle, upconversion-based spectral imaging schemes have been realized via temporal scanning of light sources \cite{Junaid2019Optica}, mechanical tuning \cite{Zhao2023NC}, or phase-matching control \cite{Junaid2018OE}. Nonlinear frequency upconversion inherently couples spectral information to propagation angle through wavelength-dependent phase matching \cite{Barh2019AOP}, suggesting a natural pathway towards snapshot spectral multiplexing without external dispersive optics. However, this intrinsic radial dispersion has so far been largely explored in the context of unmodulated imaging \cite{Fang2024NC}, limiting its applicability to simple or spectrally homogeneous scenes.

Here, we introduce snapshot MIR spectral imaging based on intrinsic nonlinear radial dispersion. In a 4$f$ upconversion imaging architecture, wavelength-dependent phase matching simultaneously performs frequency conversion and spectral multiplexing by mapping different MIR spectral components to distinct output angles, eliminating the need for external dispersive optics. By combining this intrinsic dispersion with diffuser-induced speckle encoding, spectral information is compressed and recovered without fixed coded apertures or spatial light modulators, substantially simplifying the optical architecture. Leveraging nonlinear upconversion to the visible, the approach achieves room-temperature MIR spectral imaging with sensitivity approaching the single-photon level across a broad spectral range from 2.5 to 4.0 $\mu$m. This work establishes intrinsic nonlinear radial dispersion as a fundamentally new mechanism for snapshot spectral imaging, transforming spectral encoding from an external optical function into an inherent property of the nonlinear imaging process.

\begin{figure*}[t!]
	\includegraphics[width=0.90 \textwidth]{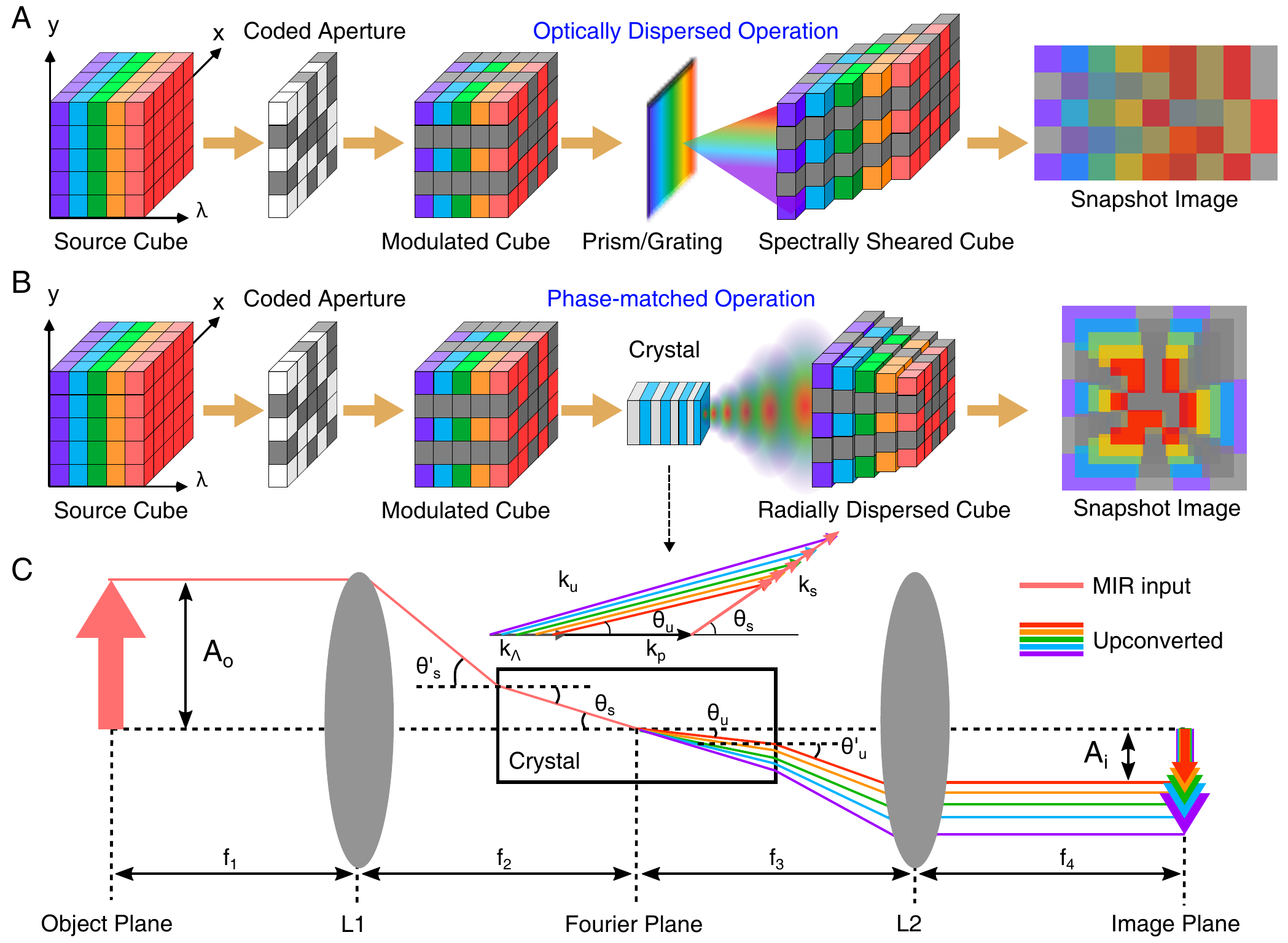}
	\caption{Data encoding and acquisition schemes for snapshot spectral imaging. (A) Principle of conventional CASSI system. Spectral images of the scene are first spatially encoded by a fixed coded aperture and subsequently linearly dispersed by a wavelength-dependent element (prism or grating) before being integrated onto a two-dimensional detector in a single exposure. (B) Concept of snapshot spectral imaging based on nonlinear upconversion and intrinsic radial dispersion. In a 4$f$ upconversion architecture, broadband MIR signals are frequency-converted in a nonlinear crystal. Wavelength-dependent phase-matching conditions map different spectral components to distinct output angles, giving rise to intrinsic nonlinear radial dispersion and enabling snapshot spectral multiplexing without external dispersive optics. (C) Schematic of parametric upconversion imaging in a 4$f$ system. Owing to wavelength-dependent phase matching, signals at different wavelengths form images with different spatial magnifications on the image plane.}
	\label{fig1}
\end{figure*}

\begin{figure*}[t!]
	\includegraphics[width=0.88\textwidth]{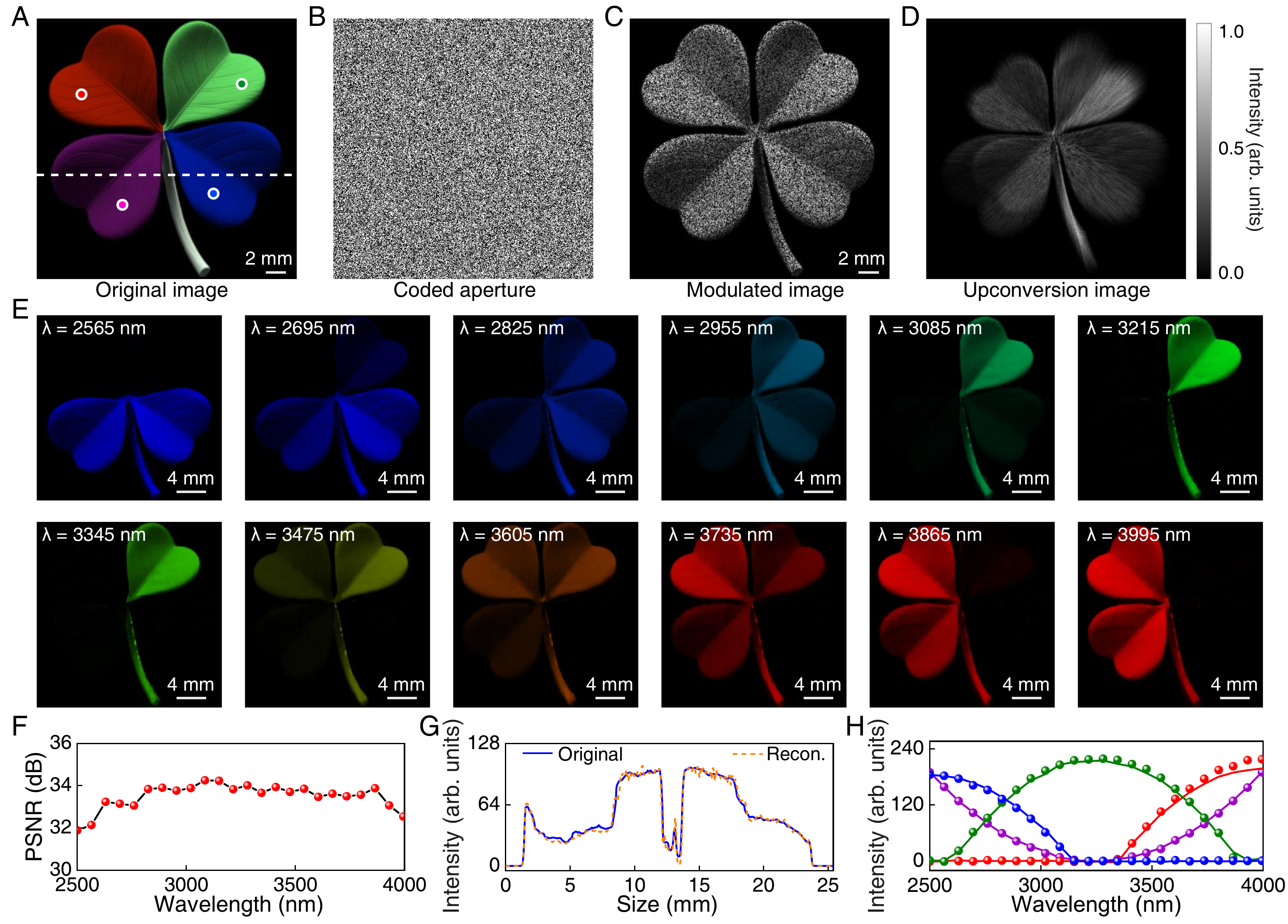}
	\caption{Numerical validation of snapshot spectral reconstruction enabled by nonlinear upconversion and radial dispersion. (A) Synthetic hyperspectral scene composed of 24 discrete spectral bands spanning 2500-3995 nm. Different leaves of the four-leaf clover carry distinct spectral signatures, visualized using RGB false-colour mapping for clarity. (B) Binary coded aperture with a spatial resolution of 512$\times$512 pixels. (C) Broadband spectral image after spatial encoding by the coded aperture. (D) Simulated snapshot image after nonlinear upconversion, exhibiting wavelength-dependent radial dispersion. (E) Representative reconstructed monochromatic images at selected wavelengths. (F) Peak signal-to-noise ratio (PSNR) of the reconstructed spectral images across all wavelengths using the GAP-TV algorithm. (G) Intensity profiles along the white dashed line in (A), comparing the reconstruction (orange dashed line) with the ground truth (blue solid line). (H) Reconstructed spectra at the marked locations in (A). Symbols denote reconstruction results, and solid lines indicate the ground truth.}
	\label{fig2}
\end{figure*}

\begin{figure*}[t!]
	\includegraphics[width=0.8 \textwidth]{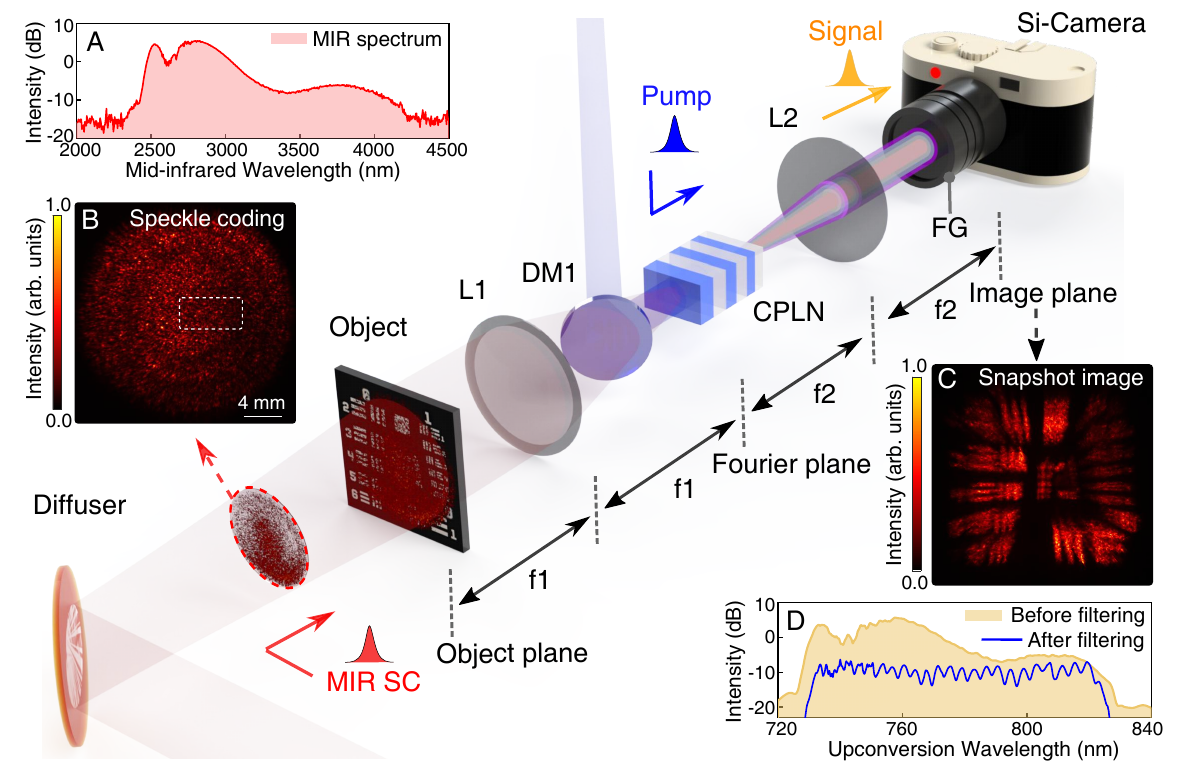}
	\caption{Experimental setup of the snapshot MIR upconversion spectral imaging system. A broadband MIR scene derived from a supercontinuum source is spatially encoded by diffuser-induced speckle patterns and relayed through a 4$f$ imaging system. A chirped-poling lithium niobate (CPLN) crystal placed at the Fourier plane enables nonlinear frequency upconversion. A synchronized 1030 nm pump beam is combined with the MIR signal to drive sum-frequency generation. The wavelength-dependent phase matching in the nonlinear crystal gives rise to an intrinsic nonlinear radial dispersion, which directly encodes spatio-spectral information into the upconverted image. Consequently, MIR spectral imaging can be achieved in a single shot. L, lens; DM, dichroic mirror; Col., collimator; FG, filter group. (A) MIR signal spectrum defining the operating wavelength range of the snapshot imaging system (2.5-4.0 $\mu$m). (B) Representative MIR speckle encoding generated by a static diffuser, shown at a central wavelength of 3073 nm. (C) Broadband upconversion snapshot image of a USAF resolution target, exhibiting wavelength-dependent radial scaling. (D) Spectrum of the upconverted signal. The yellow shaded region indicates the raw upconverted spectrum, while the blue trace shows the spectrally flattened response after shaping with an acousto-optic tunable filter.}
	\label{fig3}
\end{figure*}

\vspace{8pt}
\noindent{\fontfamily{phv}\selectfont 
	\textbf{RESULTS}}
\vspace{4pt}
\newline
\textbf{Basic principle} \\
Figure \ref{fig1}(A) presents a typical CASSI architecture, where a three-dimensional spectral datacube is spatially modulated by a coded aperture and subsequently dispersed by a wavelength-dependent element, such as a grating or prism, before being integrated onto a two-dimensional detector. Spectral encoding in this scheme relies on externally imposed linear dispersion, which produces spatial-spectral multiplexing within a single exposure. Although many pixels are blocked, this sparse structured encoding provides sufficient measurement diversity to enable accurate reconstruction of the full spectral datacube using compressive sensing \cite{Song2025OL}.

For a source scene with spectral density $f_s(x,y,\lambda)$, spatial modulation by a coded aperture with transmittance $T(x,y)$ produces a wavelength-dependent encoded scene $f_m(x,y,\lambda)=f_s(x,y,\lambda)T(x,y)$. Upon propagation through a dispersive element, each spectral component is shifted to a wavelength-dependent spatial position, resulting in a spectrally sheared image $f_i(x,y,\lambda)$ on the detector plane. The detected two-dimensional measurement is obtained by integrating these spectrally displaced contributions over wavelength,
\begin{equation}
g(x,y)=\int f_i(x,y,\lambda) \  \mathrm{d}\lambda \ ,
\label{eq1}
\end{equation}
where $f_i(x,y,\lambda)$ accounts for the encoded scene after dispersion, including the system transmission and the diffraction efficiency of the dispersive element.

In contrast, our approach replaces the external dispersive element with an intrinsic nonlinear mechanism. As illustrated in Fig. \ref{fig1}(B), nonlinear frequency upconversion in a phase-matched crystal simultaneously performs spectral translation and spectral dispersion. Energy conservation in sum-frequency generation (SFG) satisfies $\lambda_u^{-1} = \lambda_s^{-1} + \lambda_p^{-1}$, enabling indirect detection of MIR signals using silicon-based cameras \cite{Li2025Optica}. More importantly, transverse momentum conservation in the nonlinear interaction couples wavelength to propagation angle, such that different MIR spectral components emerge from the crystal with distinct output angles. This wavelength-dependent angular separation gives rise to intrinsic nonlinear radial dispersion, which naturally performs spectral multiplexing without the need for gratings or prisms.
 
The phase-matching-induced radial dispersion is quantitatively described in a 4$f$ imaging system [Fig. \ref{fig1}(C)]. Under the small-angle approximation, the wavelength-dependent imaging magnification between the object plane and the image plane is given by \cite{Fang2024NC}
\begin{equation}
M(\lambda_s)=\frac{f_2\sin\theta'_u}{f_1\sin\theta'_s}
\approx \frac{f_2\lambda_u}{f_1\lambda_s}
=\frac{f_2\lambda_p}{f_1(\lambda_s+\lambda_p)} \ ,
\label{eq2}
\end{equation}
where $f_1$ and $f_2$ are the focal lengths of the relay lenses and $\theta'_{s,u}$ denote the external signal and upconverted angles. As a result, different spectral components are radially rescaled about a common center $(x_0,y_0)$, leading to wavelength-dependent spatial mappings $x' = (x - x_0)/M(\lambda_s) + x_0$ and $y' = (y - y_0)/M(\lambda_s) + y_0$. Combining intrinsic radial dispersion with spatial encoding, the compressed measurement can be expressed as
\begin{equation}
	\begin{split}
	g(x,y) & = \int f_{s}(\frac{x - x_0}{M(\lambda_s)}+x_0, \frac{y - y_0}{M(\lambda_s)}+ y_0, \lambda_s)\eta (\lambda_s) \\
	& \ \ \ \ \ \ \ \cdot T(\frac{x - x_0}{M(\lambda_s)}+x_0,\frac{y - y_0}{M(\lambda_s)}+ y_0)\text{d}\lambda_s \ ,
\end{split}
\label{eq3}
\end{equation}
where $\eta(\lambda_s)$ includes both the system transmission and the nonlinear conversion efficiency.

For computational reconstruction, the two-dimensional measurement $g(x,y)$ and the three-dimensional spectral datacube $f_s(x,y,\lambda)$ are discretized on an $M \times N$ spatial grid with $L$ spectral channels. The measurement is vectorized as $\boldsymbol{y}\in\mathbb{R}^{M N}$, and the spectral datacube is vectorized as $\boldsymbol{x}\in\mathbb{R}^{M N L}$, yielding a linear forward model $\boldsymbol{y}=\boldsymbol{\Phi}\boldsymbol{x}$, where the sensing matrix $\boldsymbol{\Phi}\in\mathbb{R}^{MN\times MNL}$ encodes the combined effects of spatial modulation and wavelength-dependent radial dispersion. Reconstruction of the spectral datacube is then formulated as a regularized inverse problem,
\begin{equation}
	\boldsymbol{\hat{x}} = \arg \min_{\boldsymbol{x}} \frac{1}{2} \| \boldsymbol{y} - \boldsymbol{\Phi} \boldsymbol{x} \|_2^2 + \rho R(\boldsymbol{x}) \ ,
\label{eq4}
\end{equation}
where $R(\boldsymbol{x})$ encodes prior knowledge of spatial or spectral structure. In this work, we employ a generalized alternating projection algorithm with total-variation regularization (GAP-TV) \cite{Yuan2016ICIP} to demonstrate the effectiveness of nonlinear radial dispersion-based compressed acquisition. A detailed illustration of the vectorization process and the construction of the sensing matrix is provided in Supplementary Notes 1 and 2.
 
Figure \ref{fig2} presents numerical simulations validating the proposed nonlinear radial dispersion based snapshot spectral encoding and reconstruction scheme. A synthetic hyperspectral scene comprising 24 discrete spectral bands spanning 2500-3995 nm is constructed, in which different regions exhibit distinct spectral signatures [Fig. \ref{fig2}(A)]. After spatial encoding with a binary mask [Figs. \ref{fig2}(B, C)] and nonlinear upconversion, wavelength-dependent phase matching induces radial scaling of different spectral components, yielding a single snapshot in which spectral bands appear with distinct apparent spatial sizes [Fig. \ref{fig2}(D)].

This radially dispersed snapshot directly visualizes the intrinsic coupling between wavelength and spatial magnification, enabling spectral multiplexing without external dispersive optics. Using the known encoding mask and calibrated scaling factors, the full spectral datacube is accurately reconstructed. Representative monochromatic images [Fig. \ref{fig2}(E)], consistently high PSNR across all spectral channels [Fig. \ref{fig2}(F)], spatial intensity profiles [Fig. \ref{fig2}(G)] and reconstructed spectra at selected locations [Fig. \ref{fig2}(H)] all agree well with the ground truth, confirming that nonlinear radial dispersion enables accurate and invertible snapshot spectral encoding. For details of the pseudo-color rendering, see Supplementary Note 3.

\vspace{8pt}
\noindent\textbf{Imaging setup} \\
Figure \ref{fig3} shows the experimental setup of the snapshot MIR upconversion imaging system. A broadband MIR illumination is spectrally defined by a long-pass filter beyond 2.4 $\mu$m [Fig. \ref{fig3}(A)] and spatially encoded by a gold-coated diffuser [Fig. \ref{fig3}(B)]. The resulting speckle patterns provide intrinsic spatial encoding, eliminating the need for fixed coded masks or spatial light modulators and enabling a compact, alignment-tolerant architecture. The encoded object is relayed by a 4$f$ imaging system. Nonlinear frequency upconversion is performed in a chirped-poling lithium niobate (CPLN) crystal placed at the Fourier plane of the 4$f$ system, where wavelength-dependent phase matching simultaneously enables frequency conversion and intrinsic nonlinear radial dispersion. The CPLN crystal, with linearly varying poling periods from 16 to 24 $\mu$m, provides an enlarged phase-matching acceptance angle, supporting wide-field imaging and broadband spectral coverage. The resulting broadband upconverted spectrum is shown in Fig. \ref{fig3}(D).

After upconversion, the radially dispersed signal is spectrally filtered to suppress parametric fluorescence and background noise, and subsequently detected by a silicon-based electron-multiplying CCD (EMCCD) camera. The captured snapshot spectral image is shown in Fig. \ref{fig3}(C). To calibrate the wavelength-dependent speckle encoding, an acousto-optic tunable filter (AOTF) is inserted to perform narrowband filtering of the upconverted signal (see Supplementary Note 5). This procedure enables the calibration of distinct speckle encodings across MIR wavelengths spanning 2.5-4.0 $\mu$m. Furthermore, multi-frequency radio-frequency driving of the AOTF enables parallel spectral filtering and controlled spectral shaping of the upconverted signal. By independently adjusting the amplitudes of the radio-frequency components, the wavelength-dependent intensity variations of the broadband upconversion process can be effectively equalized, resulting in a flattened spectral response across channels. This spectral flattening improves the conditioning of the inverse problem and enhances the fidelity of spectral reconstruction. The resulting multi-channel upconverted spectrum is shown as the blue trace in Fig. \ref{fig3}(D). Additional experimental details are provided in Methods and Supplementary Note 4.

\begin{figure}[t!]
	\includegraphics[width=0.47 \textwidth]{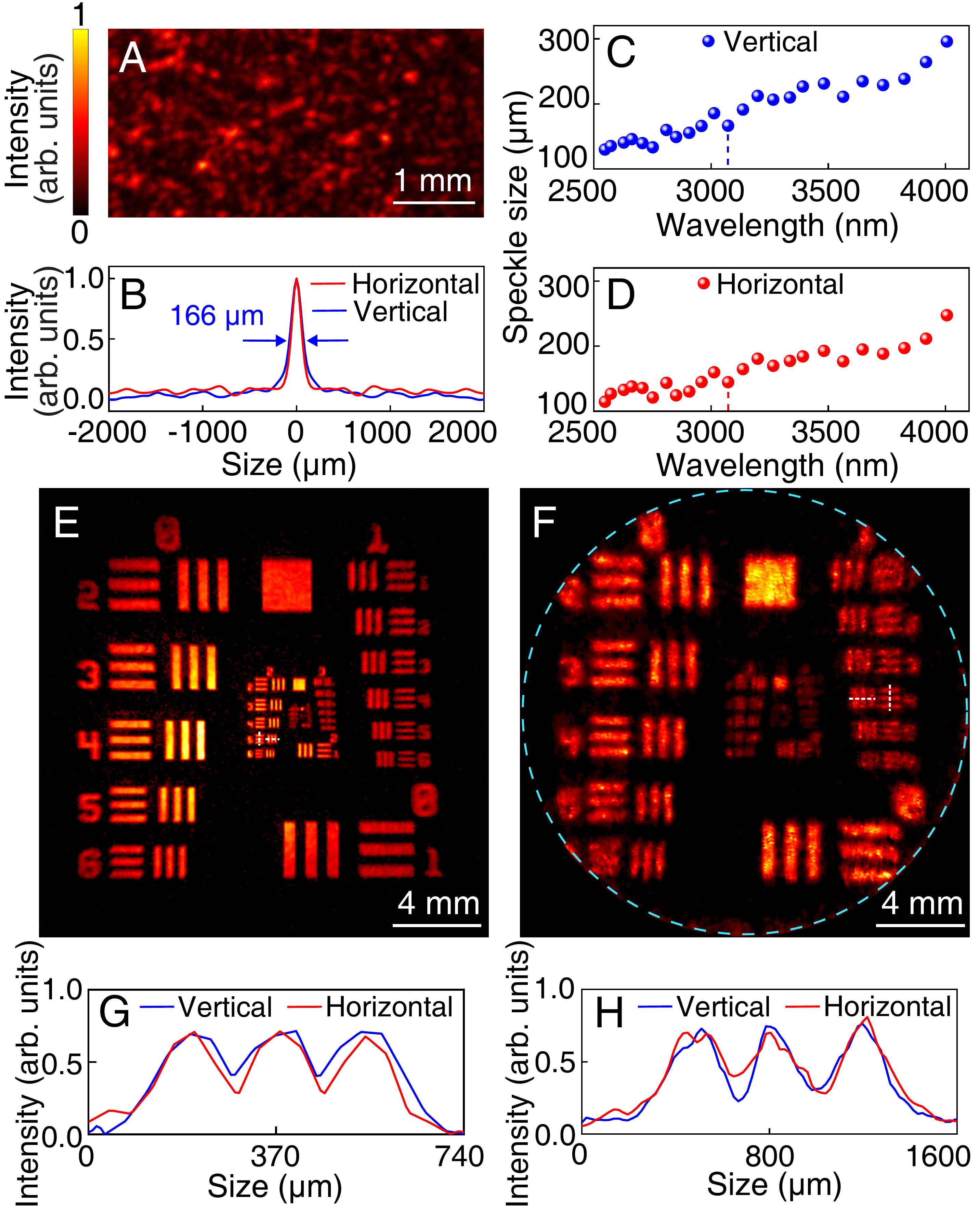}
	\caption{Spatial resolution characterization of the snapshot spectral imaging system. (A) Representative MIR speckle pattern used for spatial encoding, shown at a central wavelength of 3073 nm. (B) Normalized intensity autocovariance computed from the speckle pattern in (A). The full width at half maximum (FWHM) of the central peak, approximately 166 $\mu$m, provides an estimate of the characteristic speckle size. (C,D) Wavelength-dependent speckle sizes extracted along the vertical (C) and horizontal (D) directions, respectively. (E) Upconversion image of a USAF 1951 resolution target acquired without speckle modulation, revealing the intrinsic spatial resolution of the upconversion imaging process. (F) Reconstructed image of the same target from a single compressed snapshot measurement using speckle-based encoding. (G) Horizontal and vertical intensity profiles of the fifth element of Group 2 in (E), corresponding to the intrinsic resolution limit. (H) Horizontal and vertical intensity profiles of the fourth element of Group 1 in (F), illustrating the effective spatial resolution in snapshot spectral imaging mode.}
	\label{fig4}
\end{figure}

\begin{figure*}[t!]
	\includegraphics[width=0.75 \textwidth]{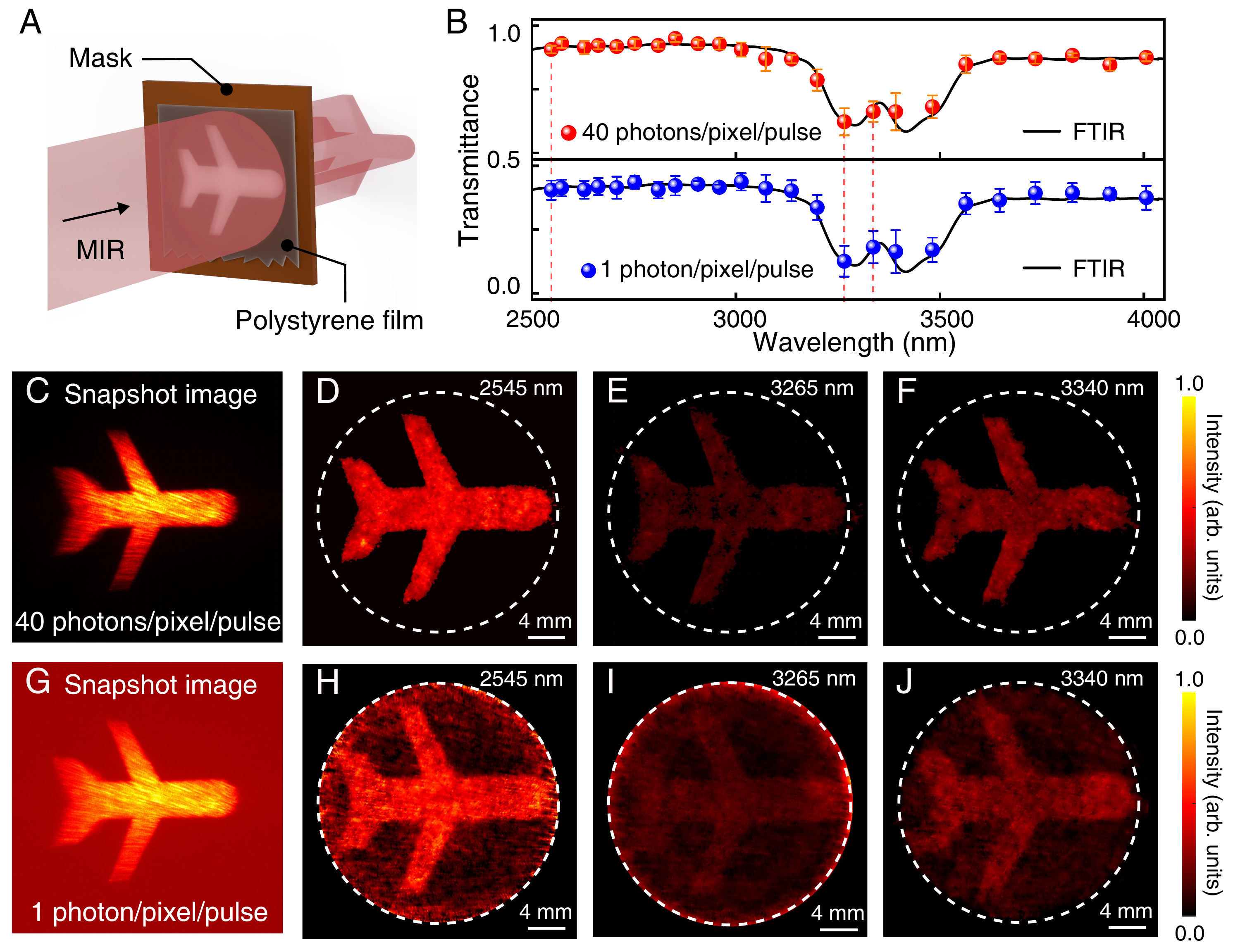}
	\caption{Single-photon snapshot spectral imaging of thin-film samples. (A) Imaging scene consisting of an aluminum plate patterned with a hollow airplane motif, partially covered by a polystyrene thin film. (B) Reconstructed spectral responses of the polystyrene film under incident photon fluxes of 40 photons/pixel/pulse (red symbols) and 1 photon/pixel/pulse (blue symbols). Note that the FTIR reference spectrum (the black curve) has been convolved to a spectral resolution of 50 cm$^{-1}$ for the sake of direct comparison. The error bars represent the standard deviation calculated from three adjacent pixels. (C) Snapshot compressed measurement acquired at 40 photons/pixel/pulse. (D-F) Reconstructed monochromatic images from (C) at central wavelengths of 2545, 3265 and 3340 nm, respectively, revealing wavelength-dependent absorption contrast. (G) Snapshot compressed measurement acquired at 1 photon/pixel/pulse. (H-J) Corresponding reconstructed spectral images from (G), demonstrating chemically specific contrast under extreme low-light conditions.}
	\label{fig5}
\end{figure*}

\vspace{8pt}
\noindent\textbf{Resolution characterization} \\
We characterize the spatial resolution of the snapshot upconversion imaging system, which is governed by two independent factors: the characteristic scale of the speckle-based spatial encoding and the intrinsic resolution limit of the nonlinear upconversion process.

The speckle encoding is calibrated in the absence of an imaging target using the AOTF. A representative speckle pattern measured at 3073 nm is shown in Fig. \ref{fig3}(B), covering a circular field of view of 20 mm in diameter. The speckle size is quantified via the normalized intensity autocovariance function [Figs. \ref{fig4}(A, B)], whose full width at half maximum provides an estimate of the average speckle width \cite{Goodman1984AP}. The speckle size is approximately 166 $\mu$m and 150 $\mu$m along the vertical and horizontal directions, respectively, and increases with wavelength, as summarized in Figs. \ref{fig4}(C, D).

The intrinsic spatial resolution of the upconversion process is limited by the spatial-frequency acceptance of the nonlinear interaction, determined by the pump beam waist and the transverse dimensions of the nonlinear crystal. The corresponding resolution limit is given by $R = 4\lambda_s f_1 / (\pi D_p)$ \cite{Barh2019AOP}. With a pump beam waist of approximately 2.5 mm, the theoretical resolution is 78.3 $\mu$m, with a slight degradation along one transverse direction due to truncation by the crystal aperture. This limit is experimentally confirmed by imaging a USAF 1951 resolution target without speckle modulation, where the fifth element of Group 2 is clearly resolved [Figs. \ref{fig4}(E, G)], yielding a measured resolution of 78.7 $\mu$m.

We then evaluate the effective resolution in snapshot spectral imaging mode. A spectral image at 3073 nm reconstructed from a single compressed measurement is shown in Fig. \ref{fig4}(F). The smallest resolvable features correspond to element 4 of Group 1, with a linewidth of 177 $\mu$m. The corresponding intensity profiles [Figs. \ref{fig4}(G,H)] exhibit contrasts of 45\% and 53\% along the horizontal and vertical directions, respectively. These results indicate that the current snapshot resolution is primarily limited by the speckle encoding rather than by the intrinsic upconversion process, suggesting a clear route toward further resolution improvement through speckle-scale engineering \cite{Yuan2021SPM}.

\begin{figure*}[t!]
	\includegraphics[width=0.9 \textwidth]{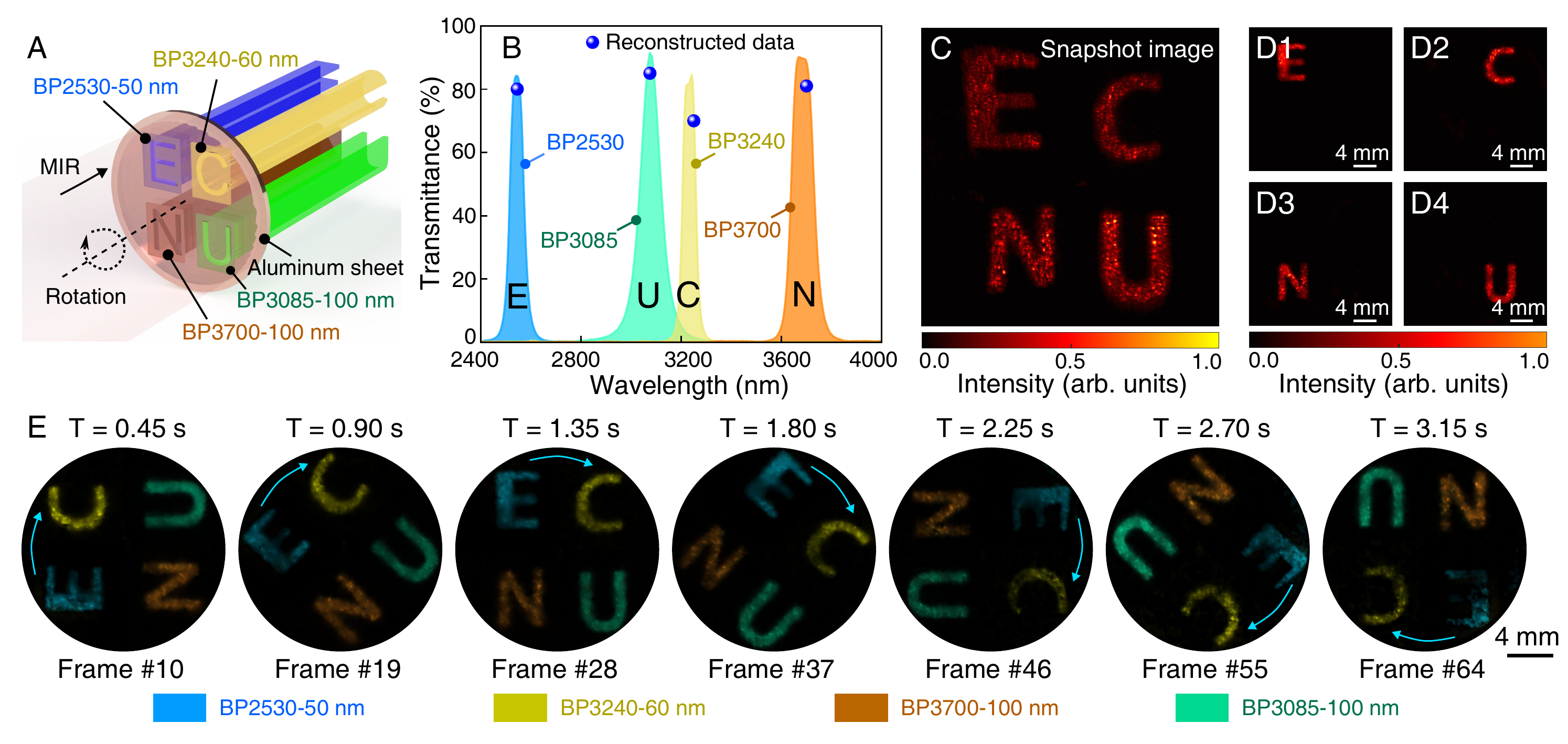}
	\caption{Real-time snapshot spectral imaging of a dynamic scene. (A) Dynamic imaging target consisting of an aluminum plate engraved with the hollow letters ``ECNU", with narrowband MIR filters of different central wavelengths placed in front of each letter. The plate and filters are rigidly mounted and rotated to emulate controlled motion. (B) Transmission bands of the narrowband filters (shaded regions) and the corresponding reconstructed spectral responses (blue symbols). (C) Raw snapshot measurement acquired in a single exposure at T = 1.35 s. (D1-D4) Reconstructed monochromatic images corresponding to the four spectral bands, revealing wavelength-dependent spatial contrast. (E) False-colour visualization obtained by combining the reconstructed spectral images into a composite RGB representation, showing the rotation dynamics captured at a frame rate of 20 Hz. A video of the dynamic snapshot spectral imaging is provided in Supplementary Video 1.}
	\label{fig6}
\end{figure*}

\vspace{8pt}
\noindent\textbf{Single-photon snapshot spectral imaging} \\
We next evaluate the performance of the MIR snapshot spectral imaging system using static scenes that simultaneously encode chemical and morphological information. The target consists of an aluminum plate etched with a hollow airplane pattern, partially covered by a polystyrene film [Fig. \ref{fig5}(A)]. Polystyrene exhibits a pronounced vibrational absorption feature near 3300 nm. A reference absorption spectrum measured by FTIR spectroscopy is shown as the black curve in Fig. \ref{fig5}(B).

Owing to the high sensitivity enabled by nonlinear upconversion, high-quality snapshot spectral images are obtained at an incident MIR power of approximately 66 pW/pixel, equivalent to a photon flux of 40 photons/pixel/pulse. A representative snapshot measurement is shown in Fig. \ref{fig5}(C). The reconstructed spectrum of the polystyrene film [red symbols in Fig. \ref{fig5}(B)] shows good agreement with the FTIR reference [black curve in Fig. \ref{fig5}(B)], where the FTIR spectrum has been convolved to a spectral resolution of about 50 cm$^{-1}$ for the sake of direct comparison. The current spectral resolution is primarily limited by the calibration bandwidth of the AOTF. Thanks to the encoding diversity introduced by the diffuser, the achieved resolution surpasses that predicted by a simple model based solely on nonlinear radial dispersion (see Supplementary Note 6). To illustrate the spatial-spectral contrast, reconstructed monochromatic images at 2545, 3265 and 3340 nm are displayed in Figs. \ref{fig5}(D-F), clearly revealing wavelength-dependent absorption while preserving morphological features. The achieved modest spectral resolution is sufficient for material identification and sorting \cite{Knez2022SA, Zhao2023NC}, as exemplified by the chemical film identification demonstration in Supplementary Note 7.

The sensitivity limit of the system is further assessed under the condition of a 5-s camera integration time by reducing the incident photon flux to approximately 1 photon/pixel/pulse (about 1.65 pW/pixel). Under this extreme low-light condition, snapshot measurements remain feasible [Fig. \ref{fig5}(G)], albeit with a reduced signal-to-noise ratio. The reconstructed spectral response, shown as blue symbols in Fig. \ref{fig5}(B), still captures the characteristic absorption feature of polystyrene. Corresponding reconstructed images [Figs. \ref{fig5}(H-J)] exhibit increased background noise but retain discernible structural and spectral contrast. The transmittance spectra are obtained by averaging over multiple pixels, with larger fluctuations observed in strongly absorbing spectral regions. All reconstructed images are spatially calibrated, with the white dashed outlines indicating the actual field of view.

\vspace{8pt}
\noindent\textbf{Real-time snapshot spectral imaging} \\
We next investigate the real-time performance of the wide-field MIR snapshot spectral imaging system. A dynamic scene is emulated by rotating a copper plate engraved with the letters ``ECNU" as shown in Fig. \ref{fig6}(A). To introduce well-defined spectral contrast, a narrowband filter with a distinct center wavelength is mounted in front of each letter. The corresponding transmission bands, centered at 2530, 3085, 3240 and 3700 nm, are indicated as shaded regions in Fig. \ref{fig6}(B).

The system operates at a frame rate of 20 Hz with an acquisition time of 50 ms per frame, limited primarily by the readout speed of the silicon camera. A representative snapshot acquired at T = 1.35 s is shown in Fig. \ref{fig6}(C). Owing to wavelength-dependent nonlinear radial dispersion, the four letters appear with distinct apparent spatial sizes within a single exposure, directly reflecting their different spectral content. The retrieved spectral responses, shown as discrete points in Fig. \ref{fig6}(B), are consistent with the known filter transmission windows. Representative reconstructed monochromatic images corresponding to each letter are shown in Figs. \ref{fig6}(D1-D4).

To visualize the spectral differentiation in a compact form, the reconstructed datacube is projected onto four spectral bands matching the transmission windows of the filters and rendered as pseudo-color channels \cite{Knez2022SA, Fang2024NC}. These channels are combined into composite images, yielding a real-time color-coded representation of the dynamic scene. Figure \ref{fig6}(E) presents a sequence of reconstructed frames, demonstrating stable spectral contrast and minimal motion-induced artifacts across consecutive snapshots. This result highlights the capability of the upconversion-based architecture to capture dynamic MIR spectral scenes at video rate, underscoring its potential for high-throughput applications such as chemical reaction monitoring and thermal process diagnostics.

\vspace{8pt}
\noindent{\fontfamily{phv}\selectfont 
	\textbf{DISCUSSION}}
\vspace{4pt}
\newline
In summary, we have demonstrated a snapshot MIR spectral imaging platform that departs fundamentally from conventional dispersive and coded-aperture architectures. By exploiting intrinsic nonlinear radial dispersion in a phase-matched upconversion process, spectral encoding is no longer imposed by external optical elements, but instead emerges as an inherent property of the nonlinear imaging interaction itself. This transformation of spectral encoding, which is from an externally engineered function to an intrinsic physical mechanism, constitutes the central conceptual advance of this work.

In contrast to traditional snapshot spectral imaging schemes \cite{Yang2023RS, Yang2021RS}, which rely on fixed coded masks and linear dispersive components to multiplex spectral information, the present approach unifies spectral encoding and frequency conversion within a single nonlinear process. Diffuser-induced speckle modulation provides flexible spatial encoding without dedicated masks or spatial light modulators, while wavelength-dependent phase matching in a chirped-poling lithium niobate crystal maps different MIR spectral components to distinct spatial magnifications. As a result, broadband spectral information is multiplexed in a single exposure through deterministic, radially dispersive scaling, enabling wide-field snapshot acquisition without sacrificing optical throughput or system compactness. 


The use of nonlinear upconversion further addresses one of the long-standing bottlenecks in MIR imaging: the lack of high-performance, room-temperature detector arrays. By translating broadband MIR signals to the visible band, the demonstrated system leverages mature silicon-based cameras to achieve high sensitivity down to the single-photon-per-pixel regime, while maintaining video-rate operation. Importantly, this sensitivity enhancement is achieved without compromising spectral bandwidth, allowing chemically specific vibrational signatures to be captured under extremely weak illumination conditions. Performance comparison among existing MIR wide-field spectral imaging techniques is summarized in Supplementary Note 8.

Beyond the specific implementation presented here, the underlying concept of intrinsic nonlinear spectral dispersion is broadly applicable. By selecting appropriate nonlinear materials and phase-matching conditions, the approach can be extended to longer-wavelength infrared and even terahertz regimes \cite{Zuo2025APL}, where efficient detectors and modulators remain scarce. Moreover, the effective spatial resolution in snapshot mode is currently limited by the speckle encoding scale rather than by the nonlinear upconversion process itself, indicating a clear pathway for further resolution enhancement through engineered scattering or wavefront modulation \cite{Blochet2025NC}. 

From a computational perspective, the demonstrated architecture naturally interfaces with modern inverse and learning-based reconstruction frameworks. Deep-learning approaches that incorporate learned spectral-spatial priors offer a promising route toward fully real-time reconstruction and visualization, further amplifying the impact of snapshot acquisition \cite{Huang2022Light}. Together, these features position intrinsic nonlinear radial dispersion as a general and versatile mechanism for high-throughput, ultrahigh-sensitivity snapshot spectral imaging, opening new opportunities for real-time studies of dynamic processes in materials science, chemistry and life sciences \cite{Shi2020NM, Xing2025Light, Yin2023SA, Ge2025PhotoniX}.

\vspace{8pt}
\noindent  {\fontfamily{phv}\selectfont 
	\normalsize \textbf{Methods}}

\noindent \textbf{\textsf{Laser sources.}} Synchronous MIR signal and pump laser sources are employed to implement coincidence pumping for nonlinear upconversion. Pulsed excitation enhances the conversion efficiency through high peak power while suppressing background noise via ultrashort temporal gating. The MIR illumination is provided by a supercontinuum fiber laser (Thorlabs, SC4500), delivering broadband radiation spanning 1.5-4.0 $\mu$m at a repetition rate of up to 50 MHz. After spectral filtering with a 2.4 $\mu$m long-pass filter, the MIR beam is modulated by a diffuser and illuminates the target with a beam diameter of approximately 1 inch, corresponding to an average irradiance of about 1.2 mW cm$^{-2}$.

The reflected component from the 2.4 $\mu$m long-pass filter, centered at 1550 nm, is injected into a nonlinear amplifying loop mirror cavity operating at 1030 nm to synchronize the pump pulses with the MIR signal via cross-phase modulation. The pump laser operates at a repetition rate of 25 MHz, corresponding to half of the MIR repetition rate. The pump pulse duration is approximately 10 ps, sufficiently long to temporally overlap the MIR pulses. After two-stage fiber amplification, the pump maintains a spectral bandwidth of approximately 1 nm with an average power of 600 mW, corresponding to a peak power of about 2.4 kW.

\vspace{8pt}
\noindent \textbf{\textsf{Speckle calibration.}} An acousto-optic tunable filter is used to calibrate the wavelength-dependent speckle encodings. The AOTF is positioned after the nonlinear crystal to spectrally filter the broadband upconverted signal, and the resulting monochromatic images are recorded by a silicon-based camera placed at the image plane of the 4$f$ system. The AOTF has an active aperture of 8$\times$8 mm$^2$, a switching time below 3 $\mu$s, and a diffraction efficiency exceeding 85\%.

To establish the correspondence between the radio-frequency drive and the diffracted wavelength, a grating-based infrared spectrometer (Simtrum, L600-900) is placed in the AOTF diffraction path to calibrate the RF-wavelength relationship (see Supplementary Note 5). For each RF frequency, the corresponding speckle pattern is recorded by the silicon-based electron-multiplying CCD camera. By adjusting the RF drive amplitude, the diffraction efficiency of the AOTF is tuned to flatten the spectral response across wavelengths, thereby improving the effective dynamic range of detection. In addition, composite RF signals comprising multiple frequency components are applied to the AOTF to enable parallel filtering of multiple spectral channels.

\vspace{8pt}
\noindent \textbf{\textsf{Reconstruction algorithm.}} Reconstruction of the spectral datacube from a single snapshot is formulated as a linear inverse problem. Because the forward model is underdetermined, additional constraints are required to obtain a unique and physically meaningful solution. In this work, we employ the generalized alternating projection with total variation regularization algorithm. The reconstruction is formulated as a TV-constrained optimization problem solved within the GAP framework, in which iterative updates are implemented as Euclidean projections. Compared with alternating direction method of multipliers (ADMM)-based approaches, GAP-TV offers reduced parameter sensitivity and lower computational complexity for spectral compressive imaging, while providing robust reconstruction quality for the present dataset. 

Reconstruction of a single 1024×1024×24 spectral datacube requires approximately 10 minutes on a standard desktop computer (Intel Core i9-14900HK, 64 GB RAM, Python 3.9, without GPU acceleration) using the GAP-TV algorithm with 100 iterations.

\vspace{8pt}
\noindent  {\fontfamily{phv}\selectfont 
\normalsize \textbf{Acknowledgments} 
}
\newline
\noindent This work was funded by Shanghai Pilot Program for Basic Research (TQ20220104); National Natural Science Foundation of China (62505088, 62235019, 125B2088); Natural Science Foundation of Chongqing (CSTB2025NSCQ-GPX0443); Shanghai Municipal Science and Technology Major Project (2019SHZDZX01); Postdoctoral Fellowship Program and China Postdoctoral Science Foundation (GZC20250545, 2024M760918, 2025T180224).

\vspace{8pt}
\noindent  {\fontfamily{phv}\selectfont 
	\normalsize \textbf{Author contributions} 
}
\newline
\noindent K. H. and H. Z. conceived the project and designed the experiments. J. F., J. Z., and K. H. built the system, performed experiments, and processed data. J. F. and R. Q. built fiber laser sources. K. H. analyzed the imaging data. J. F. and K. H. wrote the manuscript draft. All authors were involved in discussions and contributed to the manuscript editing.

\vspace{8pt}
\noindent  {\fontfamily{phv}\selectfont 
	\normalsize \textbf{Data availability} 
}
\newline
\noindent The data that support the findings of this study are available from the corresponding author upon request. Source data are provided with this paper.

\vspace{8pt}
\noindent  {\fontfamily{phv}\selectfont 
	\normalsize \textbf{Code availability} 
}
\newline
\noindent The code files underlying this study are available on GitHub: https://github.com/HKLAB-ECNU/MIR-snapshot-spectral-imaging.

\vspace{8pt}
\noindent  {\fontfamily{phv}\selectfont 
	\normalsize \textbf{Conflict of interest} 
}
\newline
\noindent The authors declare no competing interests.

\vspace{8pt}
\noindent  {\fontfamily{phv}\selectfont 
	\normalsize \textbf{Additional information} 
}
\newline
\noindent \textbf{Supplementary information} The online version contains supplementary material available at https://doi.org/XXX.

\end{document}